\documentclass[preprint,12pt,nofootinbib]{revtex4-1}

\usepackage{lineno}

\usepackage[paper=letterpaper,margin=1.5in]{geometry}

\usepackage{graphicx}\graphicspath{%
{graph/}}%
\usepackage{amssymb,amsmath,amsfonts}
\usepackage{time}
\usepackage{epstopdf}
\usepackage{hyperref}
\hypersetup{
  colorlinks,
  citecolor=magenta,
  linkcolor=blue}
\usepackage[T2A,T1]{fontenc} % UTF-8: PDFLaTeX
\usepackage[utf8]{inputenc} % UTF-8: PDFLaTeX
\usepackage{txfonts} % UTF-8: PDFLaTeX

\renewcommand{\vec}[1]{\textnormal{\boldmath$#1$}}
\begin{document}

\begin{flushright}
{\normalsize
SLAC-PUB-16803\\
%arXiv submit/1515025\\
August 2016}
\end{flushright}

\title{Measurements of Terahertz Radiation Generated
 using a Metallic, Corrugated Pipe \footnote[1]{Work supported by the U.S. Department of Energy, Office of Science, Office of Basic Energy Sciences, under Contract No. DE-AC02-76SF00515
} }

\author{Karl Bane\footnote[2]{kbane@slac.stanford.edu}, Gennady Stupakov}
\affiliation{SLAC National Accelerator Laboratory,Menlo Park, CA 94025}
\author{Sergey Antipov}
\affiliation{Euclid Techlabs LLC, Bolingbrook, IL 60440 }
\author{Mikhail Fedurin, Karl Kusche, Christina Swinson}
\affiliation{Brookhaven National Laboratory, Upton, NY 11973}
\author{Dao Xiang}
\affiliation{Department of Physics and Astronomy, Shanghai Jiao Tong University, Shanghai 200240, China }

\begin{center}
%\today
\end{center}

\begin{abstract}
A  method  for  producing  narrow-band  THz  radiation
proposes  passing   an   ultra-relativistic   beam   through   a 
metallic   pipe   with   small   periodic   corrugations.   We 
present results of a measurement of such an arrangement 
at  Brookhaven's  Accelerator  Test  Facility  (ATF).  Our 
pipe  was  copper  and  was  5  cm  long;  the  aperture  was 
cylindrically symmetric, with a 1 mm (radius) bore and a 
corrugation   depth   (peak-to-peak)   of   60   um.   In   the 
experiment  we  measured both the effect on  the beam of 
the structure wakefield and the spectral properties of the 
radiation  excited  by  the  beam.  We  began  by  injecting  a 
relatively  long  beam compared  to  the  wavelength  of  the 
radiation, but with short rise time,   to   excite   the   structure,   and   then   used   a 
downstream     spectrometer     to     infer     the     radiation 
wavelength.  This  was  followed  by  injecting  a  shorter 
bunch, and then using an interferometer (also downstream 
of  the  corrugated  pipe)  to  measure  the  spectrum  of  the 
induced THz radiation. For the THz pulse we obtain and compare with calculations: the central frequency, the bandwidth,
and the spectral power---compared to a diffraction radiation background signal.

\end{abstract}

\maketitle

\section*{Introduction}

%For applications in fields as diverse as chemical and biological imaging,  material science, telecommunication, semiconductor and superconductor research, 
There is great 
interest  in  having  a  source  of  short,  intense  pulses  of 
terahertz radiation. There are laser-based sources of such 
radiation,  capable  of  generating  few-cycle  pulses  with 
frequency over the range 0.5–6 THz and energy of up to 
100 ~$\mu$J~\cite{Lee05}.  And  there  are  beam-based  sources,  utilizing 
short,   relativistic   electron   bunches.   One   beam-based 
method impinges an electron bunch on a thin metallic foil 
and generates coherent transition radiation (CTR). Recent 
tests  of  this  method  at  the  Linac  Coherent  Light  Source 
(LCLS)  have  obtained  single-cycle  pulses  of  radiation 
that  is  broad-band,  centered  on  10  THz,  and  contains  > 
0.1   mJ   of   energy~\cite{Daranciang:11}.   Another   beam-based   method 
generates narrow-band THz radiation by passing a bunch 
through a metallic pipe coated with a thin dielectric layer~\cite{Cook09}-\cite{Antipov12}.

Another, similar method for producing narrow-band THz radiation has proposed passing the beam through a metallic pipe with small periodic corrugations~\cite{BaneStupakov_THz}, which is the subject of the present report. We consider here round geometry which will yield radially polarized THz (studies of this idea in flat geometry can also be found~\cite{Kim13}). We present results of measurements of the spectral properties of the radiation excited by the beam. 

%On February 27 and 28, 2013, a
A corrugated structure that we call ``TPIPE" was tested with beam at the Accelerator Test Facility (ATF) at Brookhaven National Laboratory. We first used a relatively long beam, with short rise time---compared to the wavelength of the radiation---to excite the structure, and then used a downstream spectrometer to infer the central wavelength of the radiation. Then for a shorter bunch, by means of an interferometer also downstream of the corrugated pipe, we measured the spectrum of the induced THz. 
%We had no way of measuring the absolute strength of excitation; however, 
Due to a background of diffraction radiation of the bunch field, we could obtain the relative strength of the THz signal to this background. Our experimental set-up was simple and not optimized for the efficient collection of the radiation (by {\it e.g.} the inclusion of tapered horns between the structure and the collecting mirror of the interferometer, as was done in Refs.~\cite{Andonian11}, \cite{Antipov12}).  As such, the present experiment should be considered a proof-of-principle experiment for generating THz using a round, corrugated, metallic structure.    

Specifically, our goal in this work is to demonstrate a narrow-band THz signal downstream of TPIPE using the two measurement methods. In addition, we measure, and compare with calculations, the central frequency of the pulse, the bandwidth, and the relative strength of the signal at the central frequency. 

\section*{Experimental Set-Up}

\begin{figure}[!tbh]
\centering
\includegraphics*[width=140mm]{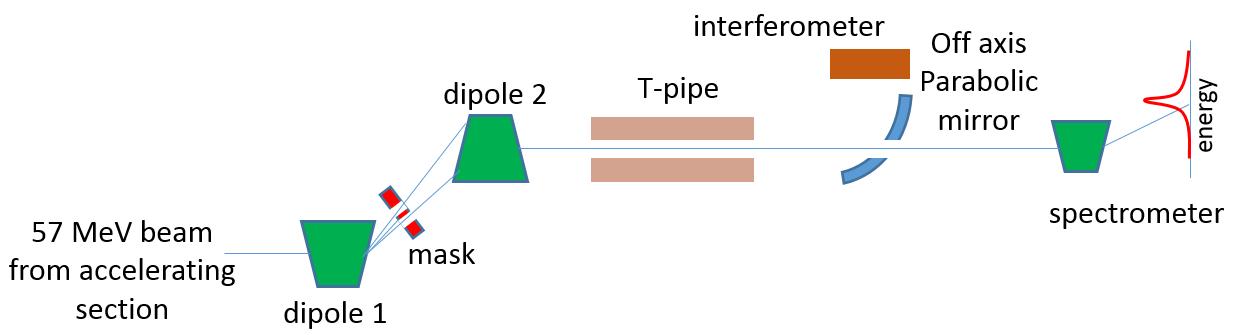}
\caption{Sketch of the experimental layout. %The profile of a section of TPIPE is given at the lower right.
}
\label{layout_Sergey_fi}
\end{figure}

A schematic of the experimental layout is presented in Fig.~\ref{layout_Sergey_fi}. A 57~MeV electron beam was initially shaped using a mask~\cite{Muggli08} and then sent through TPIPE to generate a THz pulse. The beam leaves the structure directly followed by the THz pulse. At the exit of TPIPE the radiation pulse diffracts, and further downstream some of it is reflected by an off-axis parabolic mirror into a Michelson interferometer for characterization. The intensity of the THz signal collected is measured by an IRLabs General Purpose 4.2K (liquid helium) Bolometer System. Note that the mirror is located 17.5~cm beyond TPIPE, and that in front of the mirror is a 12.5~mm radius iris, which limits the radiation that is collected. As for the electron beam, it passes through a 2.5~mm radius hole in the mirror, where it generates diffraction radiation, some of which also ends up in the interferometer. Finally, the electron bunch enters the spectrometer for characterization. 

To shape the beam we started by accelerating it off-crest in the accelerating section, in order to create a linear correlation between the longitudinal bunch coordinates $z$ and energy $E$. When the beam passes through a dipole magnet (``dipole 1" in Fig.~\ref{layout_Sergey_fi}) it becomes horizontally dispersed as in an energy spectrometer. A transverse mask is placed after the first dipole to block electrons of certain energies. A second dipole of opposing sign (``dipole 2") restores the beam to its original state, minus, however, the electrons that were blocked. %However, due to the original $E$-$z$ correlation in the beam, the result is that the mask shape is left imprinted on the longitudinal charge distribution. In the experiment we monitored the shape of the beam immediately after the mask and were able to calibrate the image for the purpose of measuring the bunch length. This was done in the following way (with TPIPE kept retracted from the beam path): A mask with three periodic holes was inserted into the beam's path. The resulting three bunches excited a periodic signal by means of diffraction radiation on passing through the hole in the parabolic mirror. We measured the periodicity of the excited signal using the interferometer.  Thus we were able to relate distances in pixels on the beam monitor image (after the mask) to longitudinal dimensions in microns on the shaped beam. 
Due to the $E$-$z$ correlation of the beam, the image in the downstream spectrometer also carries information about the beam's longitudinal shape. 
%In another experiment [Antipov] under the same conditions, a longitudinally arrow-shaped beam maintained the arrow shape in the spectrometer image. 
Therefore, distances on the spectrometer image can be related to longitudinal distances within the beam. 
%In this argument we have ignored non-linear distortions to the beam's phase space, such as can be caused by coherent synchrotron radiation in the dipoles. 
%From simulations with MAD, we estimate that such effects impose at most a 5\% error on our bunch length (?) measurements [H. Grote, C. Iselin, The MAD program (methodical accelerator design): version 8.10; user reference manual (Geneva, CERN, 1993)] simulations. Similarly to [Antipov] a beam longer than the TPIPE wavelength picked up an energy modulation due to periodic self-wakefield (see wakefield plot in the theory part figure) when traveling through TPIPE. The TPIPE mount was motorized and could be moved in and out of the beam. TPIPE also has a smooth bore with the same inner diameter (2~mm) to be used for reference scans. 
%We estimate that such effects are small.

\section*{Calculations}

\subsection*{Central Frequency of THz Pulse}

\begin{figure}[!tbh]
\centering
\includegraphics*[width=60mm]{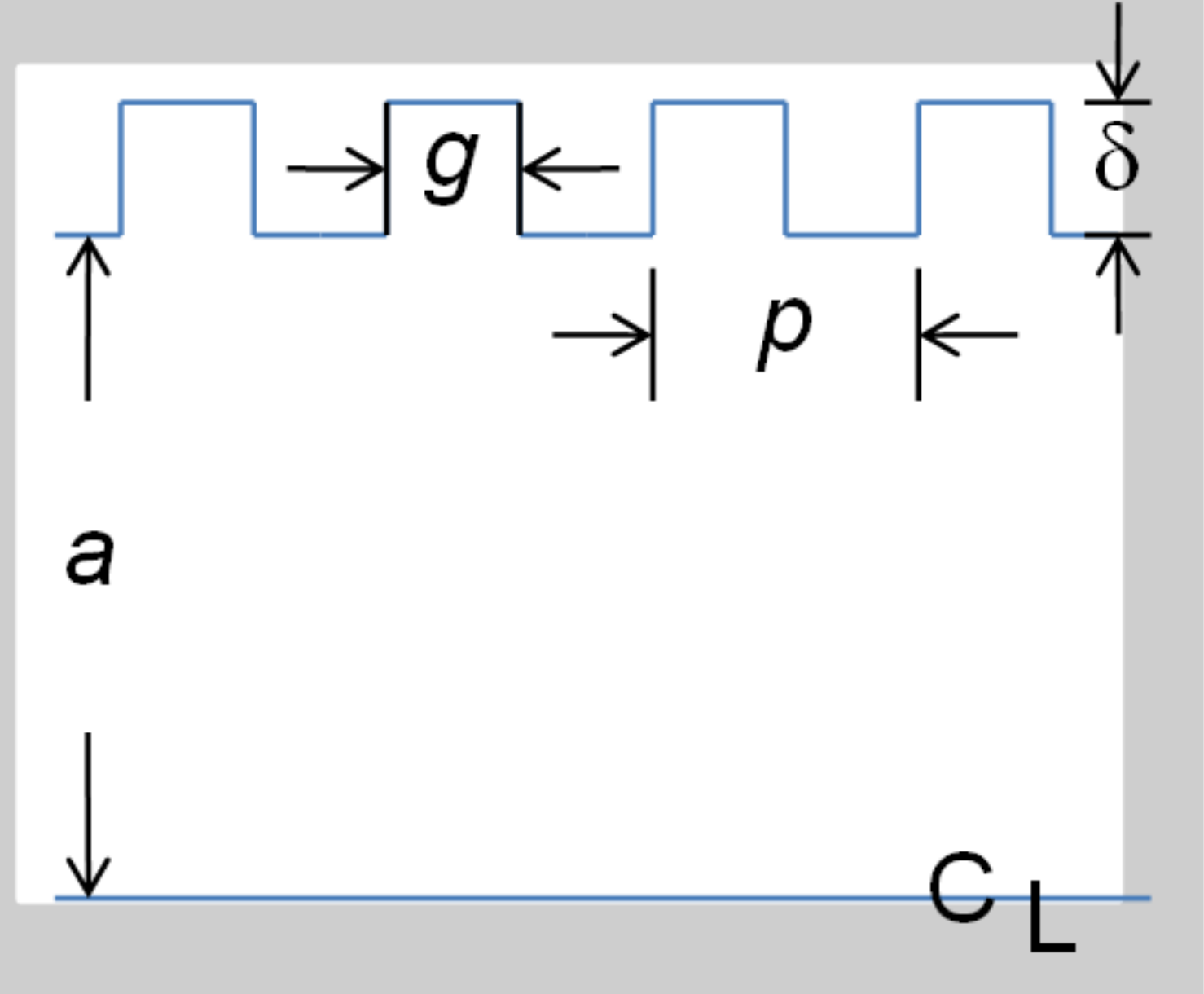}
\caption{Idealized geometry of the corrugated structure.  ``CL" denotes the axis of the pipe.}
\label{geometryr_fi}
\end{figure}

Consider a metallic beam pipe with a round bore and small, rectangular (in longitudinal view) corrugations (see Fig.~\ref{geometryr_fi}). The parameters are period $p$, (full) depth of corrugation $\delta$, corrugation gap $g$, and pipe radius $a$; where we consider small corrugations ($\delta$, $p$) $\ll a$ and also $\delta\gtrsim p$.  Let us here assume $p=2g$. It can be shown~\cite{BaneStupakov_THz} that a short, relativistic bunch, on passing through such a structure, will induce a wakefield that is composed of one dominant mode, of wave number $k\approx2/\sqrt{a\delta}$, relative group velocity $v_g/c\approx1-2\delta/a$, with $c$ the speed of light, and loss factor $\kappa\approx Z_0c/(2\pi a^2)$, with $Z_0=377$~$\Omega$. In addition to the effect on the beam, a radiation pulse of the same frequency, with a uniform envelope (with a relatively sharp rise and fall) of full length $\ell=2\delta L/a$ ($L$ is pipe length) will follow the beam out the downstream end of the structure. One can see that, in order to generate a pulse of frequency $\sim1$~THz, both the bore radius and the corrugation dimensions must be small; with $a\sim1$~mm, then $\delta\lesssim10$~$\mu$m. For $a=1$~mm, $\delta=60$~$\mu$m, the pulse frequency $f\approx0.4$~THz. If, in addition, the pipe length is $L=5$~cm, then the full radiation pulse length, $\ell=6$~mm. 

\begin{figure}[!tbh]
\centering
\includegraphics*[width=140mm]{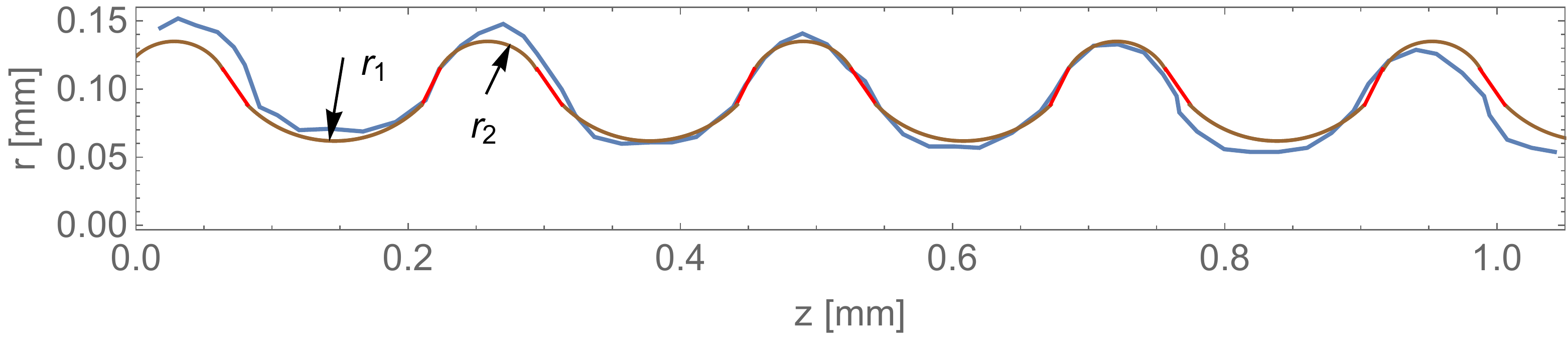}
\caption{A rough digitization of the measured geometry of TPIPE's corrugations (blue curve). The model used in ECHO simulations is also shown: arcs (brown curves) of radii $r_1=91$~$\mu$m and $r_2=41$~$\mu$m are connected by straight lines (in red), to give a structure with a period $p=231$~$\mu$m and a depth of corrugation of 72~$\mu$m.}
\label{geometry_fi}
\end{figure}

TPIPE was machined from two rectangular blocks of high purity copper, each of dimension 2 cm by 1 cm by 5 cm on a side. Two 1-mm (radius) cylindrical grooves were first machined in the long direction in each block. One groove in each block was meant to remain smooth, for the null test of the experiment. The other groove was further machined, to give it corrugations. Originally, the goal was to have rectangular corrugations (in longitudinal view) with period $\sim250$~$\mu$m and (total) depth of corrugation  $\sim60$~$\mu$m. However, rectangular corrugations of such small size are difficult to make; thus, a rounded profile was obtained using a 50~$\mu$m (radius) milling bit. When the machined blocks were measured at SLAC, it was determined that the corrugations actually had a period of $231$~$\mu$m and a depth of corrugation of 72~$\mu$m; furthermore,
the radii of the ``irises" and the ``cavities" (in longitudinal view) were not identical, and were $r_1=91$~$\mu$m and $r_2=41$~$\mu$m, respectively (see Fig.~\ref{geometry_fi}, the blue curve). As the final step, the two copper blocks were diffusion bonded at SLAC to yield one block with two cylindrically symmetric bores, one corrugated and one smooth.

We performed time-domain simulations with the 2D Maxwell equation solving program ECHO~\cite{ECHO}. We used a Gaussian bunch, with $\sigma_z=90$~$\mu$m, and let it pass through an entire 5-cm-long TPIPE structure. 
%In Fig.~\ref{geometry_fi} we show a rough digitization of the measurement of TPIPE's corrugation geometry (blue curve). 
For the model used in the ECHO simulations, we used circular arcs (brown curves in Fig.~\ref{geometry_fi}) of radius $r_1=91$~$\mu$m and angular extent $\theta=\pi/2$ (for the irises), and of radius $r_2=41$~$\mu$m and angular extent $\theta=2\pi/3$ (for the cavities), and connected them by straight lines (in red). The final structure has period $p=231$~$\mu$m and a depth of corrugation of 72~$\mu$m. We see that the model gives a good approximation to the measured shape of TPIPE (the blue curve).

\begin{figure}[!tbh]
\centering
\includegraphics*[width=94mm]{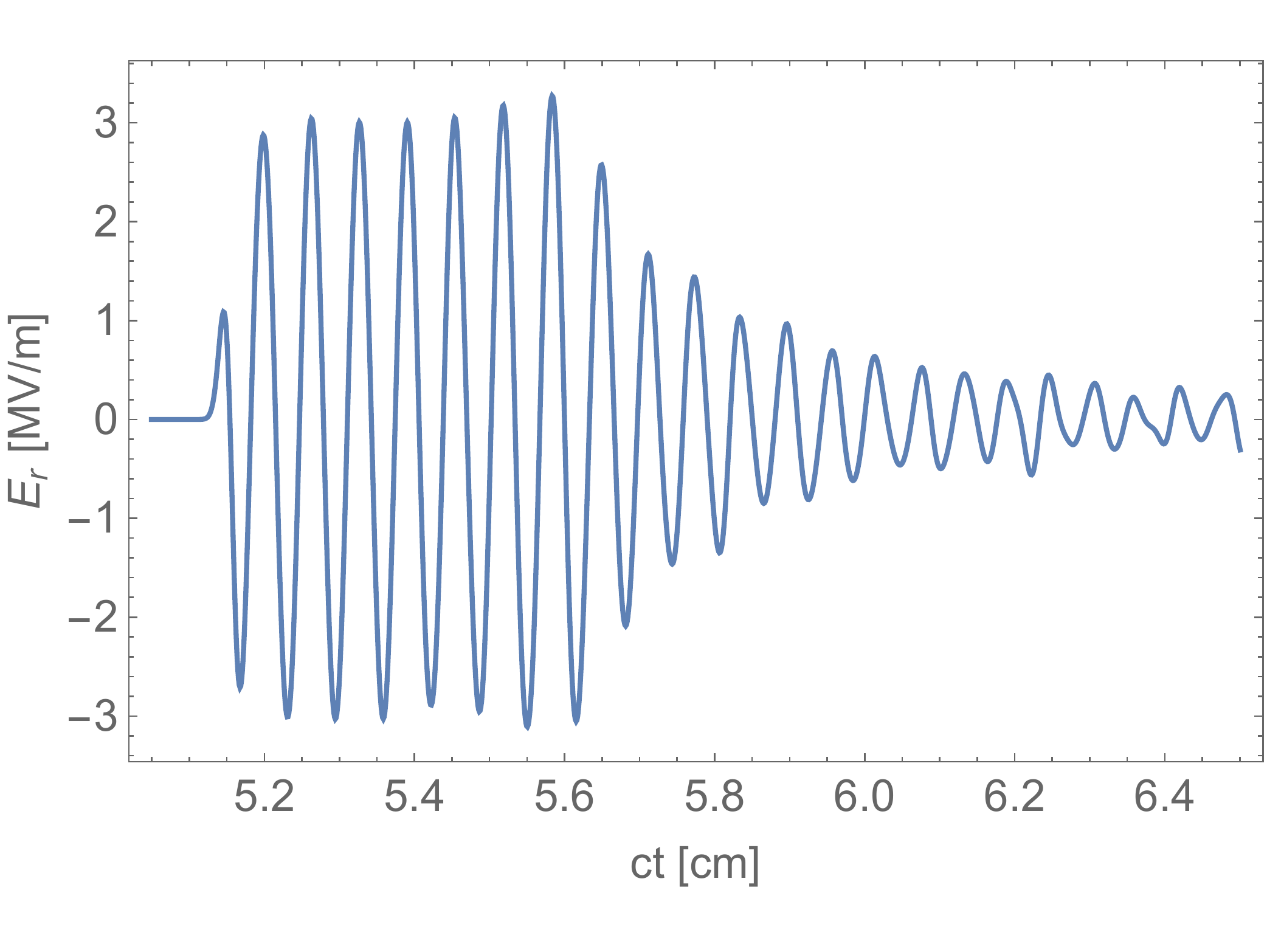}
\includegraphics*[width=94mm]{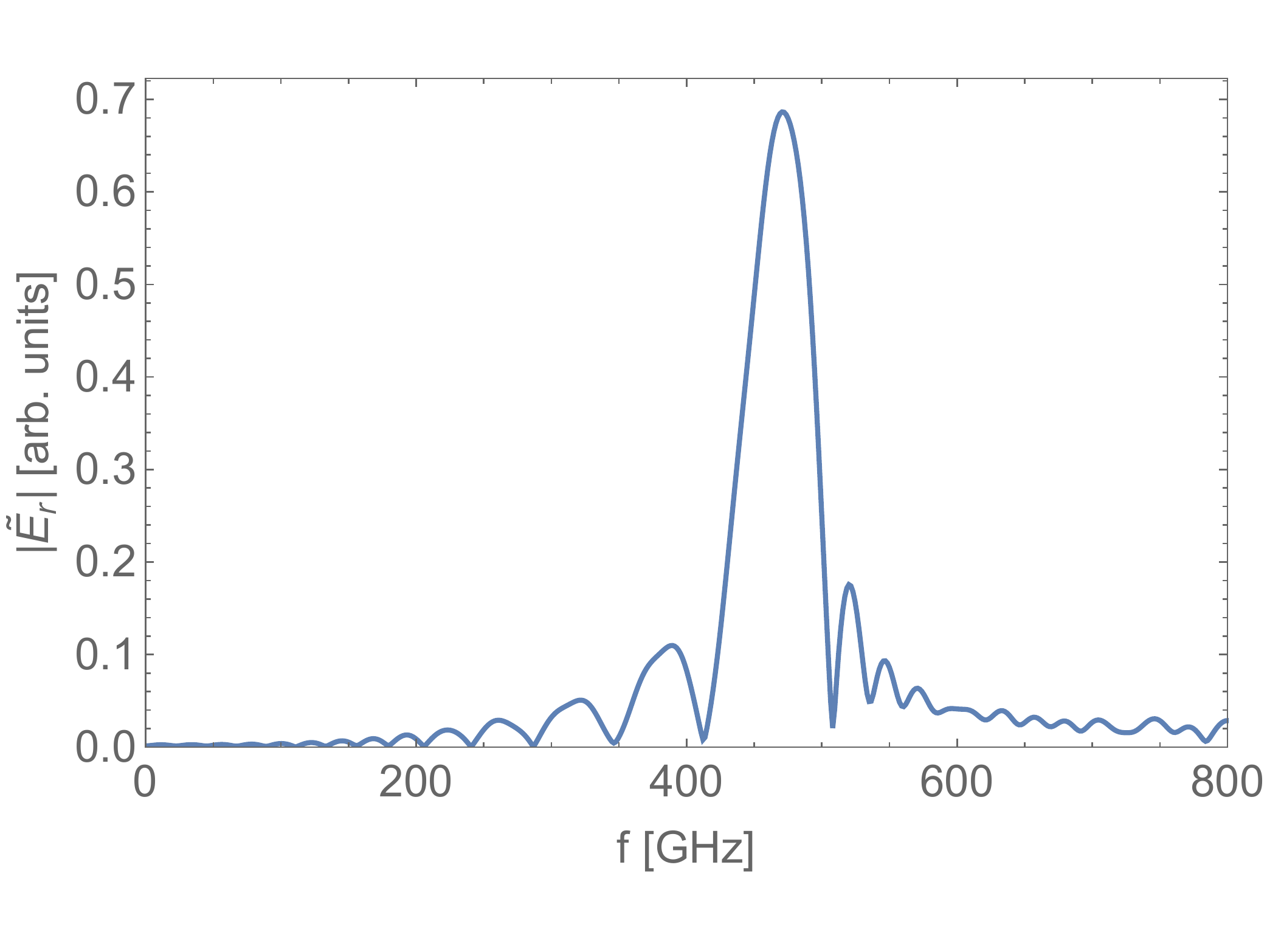}
\caption{ECHO simulation results: The radial electric field $E_r$ at radius $r=0.75$~mm at a monitor located at the downstream end of TPIPE as function of time $t$ (top); here the driving bunch is Gaussian with $\sigma_z=90$~$\mu$m and charge $q=50$~pC. The absolute value of the Fourier transform of this function, $\tilde E_r$, as function of frequency $f$, is displayed at the bottom.}
\label{Er_fi}
\end{figure}

ECHO can calculate the wakefields that a Gaussian bunch excites when it passes through a vacuum chamber object. It can also monitor the electromagnetic fields that are excited as a function of time at a given location. In our simulations we monitored the fields at a fixed $z$ location at the downstream end of our model of TPIPE. In Fig.~\ref{Er_fi} (at the top) we show $E_r(t)$ ($t$ is time) at the monitor at $r=0.75$~mm. The bunch enters TPIPE (beginning at location $z=0$) at time $t=0$; it passes the monitor at $ct=z=5.15$~cm at the end of the structure. We see that the pulse begins with a relatively fast rise time followed by an oscillation with a uniform envelope. At the trailing end, however, there is a rather long tail.
%, which is unexpected and not understood.  
(In the same kind of calculation, for a different but similar corrugated structure, the tail is much less pronounced---see Fig.~5 in~\cite{BaneStupakov_THz}.) In the process of its generation, the THz pulse ends up long compared to the driving bunch because its group velocity ($v_g<c$) is less than the bunch velocity ($v\approx c$)~\cite{BaneStupakov_THz}; the long tail of the THz pulse was thus generated near the entrance of the structure, and is apparently an initial transient effect.
When it leaves the structure, the THz pulse has become $\ell\sim5.8$~mm long (FWHM of envelope), consisting of $\sim9$ oscillations.

%The result is that a 6.5-mm-long THz pulse is generated, with frequency $f=403$~GHz and quality factor $Q=12$.  As driving bunch, we have used a Gaussian, with rms length $\sigma_z=50$~$\mu$m and charge $Q=50$~pC. 

Performing the Fourier transform of $E_r$ and taking the absolute value, we obtain the spectrum of $|\tilde E_r|$, shown in Fig.~\ref{Er_fi} at the bottom. We see a clear, relatively narrow frequency spike, with central frequency $f_c=471$~GHz. For a $q=50$~pC bunch, the average energy loss is 13~keV (neglecting the loss at the entrance and exit planes of TPIPE), 90\% of which (or 0.57~$\mu$J) ends up in the THz pulse, and 10\% in Joule heating of the structure walls. According to ECHO, the spectral energy density (at the exit of TPIPE) at the central frequency is $(dU/df)_{\mathrm{T}}=0.0125$~$\mu$J/GHz. 

%%%%%%%%%%%%%%%%%%%%%%%%%%%%%%%%%%%%%%%%%%%%%%%

\subsection*{Bandwidth of THz Pulse}

The spectrum width of the THz pulse can be characterized by the quality factor $Q$ or, equivalently, relative bandwidth $\delta f$, where $Q\equiv1/\delta f=f_c/\Delta f=8.3$ ($\Delta f$ is the full-width-at-half-maximum [FWHM] of the spectrum). (Note that the relative bandwidth {\it in energy} is approximately given by $\delta f/\sqrt{2}$.) There are two sources of the finite bandwidth of the pulse: a frequency variation within the pulse and the finite length of the pulse. 
%The total effect can be approximated by 
%\begin{equation}
%\frac{1}{Q^2}=\frac{1}{Q_{\delta f}^2}+\frac{1}{Q_{\ell}^2}\ ,\label{Q_eq}
%\end{equation}
%with $Q_{\delta f}$ ($Q_{\ell}$) the effective quality factor due to frequency variation (finite pulse length).

\begin{figure}[htb]
\centering
\includegraphics*[width=71mm]{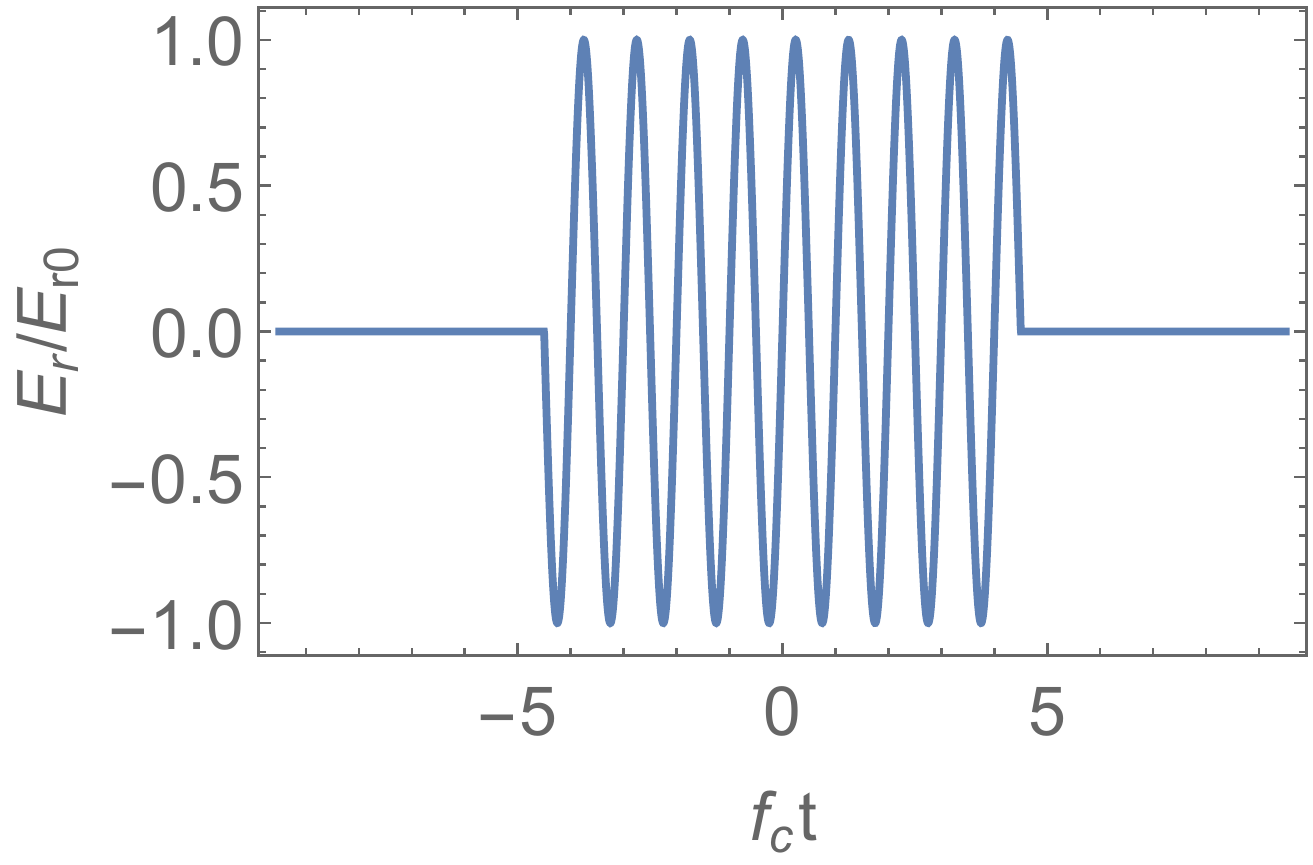}
\includegraphics*[width=67mm]{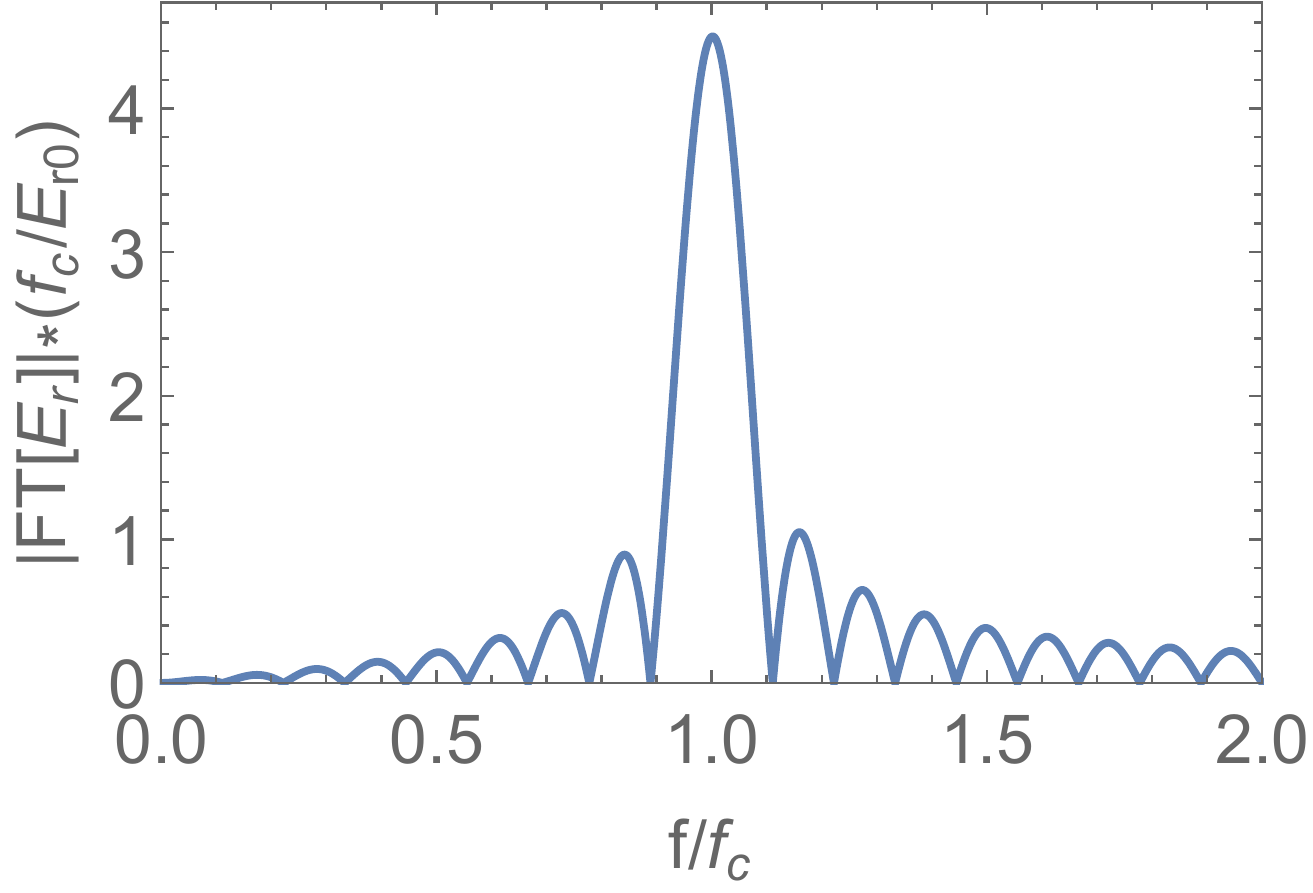}
\caption{(Left) Idealized (model) pulse from a dechirper; (right) absolute value of Fourier transform of model pulse. The pulse contains $n_\lambda=9$ wavelengths; the resulting quality factor $Q=7.5\approx n_\lambda/1.2$.}
\label{model_pulse_fi}
\end{figure}

To estimate the effect of the finite pulse length we can start with an idealized model for the pulse field:
\begin{equation}
E_r(t)=E_{r0}H(t_0-t)H(t_0+t)\cos(2\pi f_ct)\ ,
\end{equation}
where the peak field is $E_{r0}$, the unit step function, $H(x)=1$ (0) for $x>0$ ($<0$), a fixed frequency is $f_c$, and the parameter $t_0=n_\lambda/2f_c$, with $n_\lambda$ the number of oscillations in the pulse (an integer or half integer). 
Defining the Fourier transform by $\tilde E_r(f)=\int_{-\infty}^\infty E_r(t)e^{2\pi i f t}dt$, we obtain
\begin{equation}
|\tilde E_r(f)|=\frac{E_{r0}}{\pi}\frac{f}{f^2-f_c^2}
\left\{\begin{array}{r@{\quad:\quad}l}
\sin( n_\lambda\pi f/f_c)&n_\lambda{\rm\  integer}\\
\cos( n_\lambda\pi f/f_c)&n_\lambda{\rm\  half\ integer}\end{array}\right.\ .\label{Etilde_rect_eq}
\end{equation}
An idealized example of the pulse field, with $n_\lambda=9$, is shown in Fig.~\ref{model_pulse_fi} (the left frame). In the right frame we display $|\tilde E_r(f)|$. The frequency of the peak divided by its FWHM defines the quality factor, $Q$. We find that, for this simple model, when $n_\lambda\gtrsim3$, then $Q\approx n_\lambda/1.2$, which here equals 7.5.

Let us return to the ECHO calculation for TPIPE, where the $Q$ is obtained from the width (FWHM) of the spectrum peak in Fig.~\ref{Er_fi} (bottom plot), {\it i.e.} $Q=8.3$. 
%Meanwhile, the above simple model, using pulse length $\ell=5.8$~mm, $f_c=471$~GHz, yields an effective number of oscillations $n_\lambda=9$ and quality factor $Q_\ell=7.5$. 
From analyzing the ECHO data (Fig.~\ref{Er_fi}, the top plot) we find 
%we estimate that the rms frequency variation within the pulse gives an effective $Q_{\delta f}\sim35$. 
%From Eq.~\ref{Q_eq} we see that  this implies 
%To calculate the frequency variation within the pulse, we can compare the spacing of the peaks of the field obtained by ECHO  (Fig.~\ref{Er_fi}, the top plot) with an oscillation at a constant $f_c=471$~GHz; we find the weighted (relative) rms deviation to be $0.05$, implying 
that the effect on bandwidth of the frequency variation within the pulse is negligible compared to the effect of the finite length of the pulse. Thus, a measurement of the bandwidth can give us a reasonable estimate of the THz radiation pulse length
\begin{equation}
\ell=cn_\lambda/f_c\approx1.2cQ/f_c\ .\label{ell_eq}
\end{equation} 
%with $\lambda_c$ the wavelength of the central frequency.
With $Q=8.3$, this formula yields, $n_\lambda=10$ and $\ell=6.4$~mm, in reasonable agreement with the ECHO result of Fig.~\ref{Er_fi} (the top plot).

\subsection*{Relative Intensity of Signals Entering the Interferometer}

When the beam leaves TPIPE, it is directly followed by the THz pulse. The THz pulse diffracts at the exit of TPIPE and propagates downstream, where some of it is intercepted by the mirror and enters the interferometer. As for the beam, as it reaches the mirror and passes through a hole in it, it generates a diffraction radiation pulse---sometimes called a coherent transition radiation (COTR) pulse; some of this pulse is also captured by the interferometer. Here we endeavor to calculate: (1)~the fraction of THz pulse that reaches the mirror and (2)~the amount of diffraction radiation that is generated by the beam at the mirror.

Note that if there are two sources of radiation, their fields or impedances just add, from the principle of linear superposition of fields. Their power spectra, however, can interfere. We assume here that the power spectra of the THz signal and diffraction radiation background just add. This is true in our case, to good approximation, since the two sources of radiation are (mostly) separated in time: the THz pulse follows right behind the bunch, where the full bunch length is $\sim2\sigma_z\sim 200$~$\mu$m, whereas the THz pulse length is $\sim6$~mm long.

As mentioned above, ECHO found that, at the exit of TPIPE, the spectral power in the THz pulse, at the central frequency (471~GHz), is $(dU/df)_{\mathrm{T}}=0.0125$~$\mu$J/GHz. The THz pulse diffracts when it leaves TPIPE. In Appendix~A we derive an expression that can be used to find the fraction of spectral power in the pulse that reaches the mirror downstream~(see Eq.~\ref{e,q:9}):
    \begin{align}
     \rho_{\mathrm{THz}}
    =
    F\left(\frac{kab}{L}\right)
   \ ,
    \end{align}
    \begin{align}
    F(x)
    =
    1
    -
    \left(\frac{4}{x^2}+2\right) J_1(x){}^2
    -2
    J_0(x){}^2
    +
    \frac{4 }{x}J_1(x)
    J_0(x)\ ,
    \end{align}
with $k=2\pi f/c$ the wave number, $a$ the radius of the corrugated pipe, $b$ the radius of the mirror, $L$ the distance between the end of the corrugated pipe and the mirror, and $J_0(x)$, $J_1(x)$, are Bessel functions of the first kind.

For the parameters of the experiment (see Table~\ref{parameters_tab}) and central frequency (according to ECHO) $f=ck/2\pi = 471$ GHz, we find that the fraction of THz pulse energy that reaches the mirror is
    \begin{align}
    \rho_{\mathrm{THz}}
    =
    3.6\times 10^{-3}
    \ .
    \end{align}

\begin{table}[h!]
\centering
\caption{Parameters used in the analysis of the interferometry measurements.}
\begin{tabular}{||l|c|c||} \hline\hline
Parameter&Value & Unit\\ \hline\hline 
Energy, $E$ & 57.& MeV\\
TPIPE bore radius, $a$ & 1.0 & mm\\
Distance to mirror, $L$ &17.5 & cm\\
Mirror radius, $b$ &12.5 & mm\\
Mirror hole radius, $b_1$&2.5 & mm\\ \hline\hline
\end{tabular}\label{parameters_tab}
\end{table}

In Appendix B we derive the spectral power in the diffraction radiation pulse, generated when the beam passes through the hole in the mirror. The result is (\ref{eq:17})
    \begin{align}
   \left( \frac{dU}{df}\right)_{\mathrm d}
    =
    \frac{4q^{2}|\tilde \lambda(k)|^{2}}{c}
    \left[
	G\left(\frac{kb}{\gamma}\right)
	-
	G\left(\frac{kb_{1}}{\gamma}\right)
	\right]
   \  ,\label{Udiff_eq}
    \end{align}
    \begin{align}
    G(x)
    =
    \frac{1}{2} x^2
   \left[K_1(x){}^2-K_0(x)
   K_2(x)\right]
   \ ,
    \end{align}
with $q$ bunch charge, $\tilde \lambda(k)$ the Fourier Transform of the bunch distribution, $b$ radius of the mirror, $b_1$ radius of the hole in the mirror, $\gamma$ Lorentz energy factor, and $K_0(x)$, $K_1(x)$, modified Bessel functions of the second kind.
For a Gaussian bunch $| \tilde \lambda(k)|^{2}=e^{-k^{2}\sigma_{z}^{2}}$.
Substituting into Eq.~\ref{Udiff_eq} the parameters of Table~\ref{parameters_tab} and: $q=50$~pC,   $k=2\pi f_c/c=9.9$~mm$^{-1}$ ($f_c=471$~GHz), $\gamma = 120$, and $\sigma_{z} = 90$~$\mu$m, we obtain
    \begin{align}\label{eq:20}
     \left( \frac{dU}{df}\right)_{\mathrm d}
    =
    1.5\times 10^{-4}
    \,\,\mu\textrm{J/GHz}
    .
    \end{align}
 
Finally, at the mirror, at frequency $f=471$~GHz, the ratio of the spectral power in the THz pulse to that in the diffraction radiation pulse is
\begin{equation}
\Lambda\equiv  \rho_{\mathrm{THz}}\left( \frac{dU}{df}\right)_{\mathrm T}\Bigg/\left( \frac{dU}{df}\right)_{\mathrm d}=0.30\ .
\end{equation}
Note that, because the comparison of the spectral power of the two sources is done at one frequency, where the excitation of both signal and background depend on $| \tilde \lambda(k)|^{2}$,  the final result is relatively insensitive to bunch length.

Below we will compare these results with interferometer measurements. We have made simplifying assumptions. We have calculated the relative spectral power at the mirror, assuming the mirror is flat and perpendicular to the beam trajectory; in reality the mirror is parabolic and tilted at an angle. We have also assumed that the THz signal and diffraction radiation background add in the energy spectrum, but there will be some interference---though we believe it to be small. %Finally, we have assumed that both signal and background will pass with equal efficiency through the interferometer to the bolometer, which may not be true. 

%%%%%%%%%%%%%%%%%%%%%%%%%%%%%%%%%%%%%%%%%%%%%%%%%

\section*{Results}

\subsection*{Spectrometer Measurements}

\begin{figure}[htb]
\centering
\includegraphics*[width=120mm]{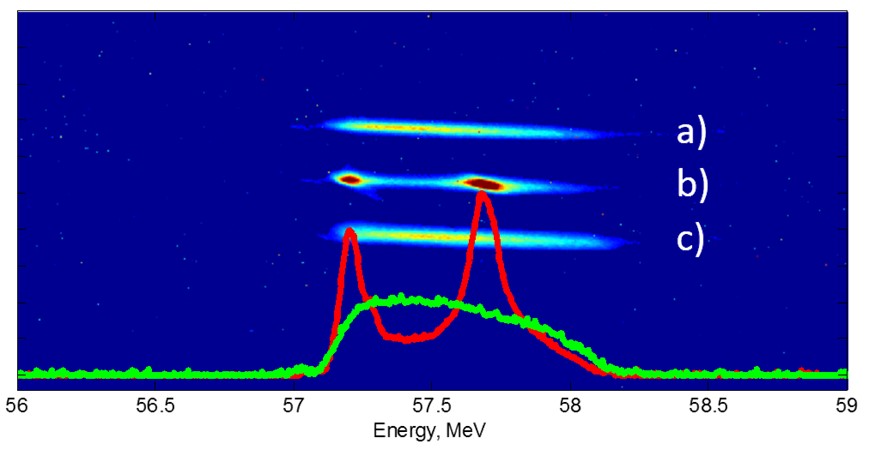}
\caption{Spectrometer images: no structure (a), TPIPE (b), smooth pipe (c). In all cases the beam is initially chirped, resulting in a horizontal streak on the screen. The green line gives the energy distribution of case (a), the red line that of case (b).}
\label{spect_fi}
\end{figure}

If we pass a long bunch with a short rise time---both compared to TPIPE's wavelength---through the structure, the beam will become energy modulated at the structure frequency (see {\it e.g.} \cite{Antipov13}). If the long bunch has an energy chirp, then the wavelength of the mode can be measured on the spectrometer screen.
In Fig.~\ref{spect_fi} we compare the energy spectrum measurements of: the original beam, with TPIPE removed from its path (a), the beam passing through the corrugated pipe in TPIPE (b), and the beam passing through the smooth tube in TPIPE (c). The green line gives the energy distribution of case (a), the red line that of case (b). First, we note that the results for the cases of no TPIPE and the smooth pipe are similar, showing a chirped streak with no modulation. With the corrugated pipe (b) we see a clear modulation. The bunch length was measured by coherent transition radiation interferometry~\cite{Muggli08}. Based on the spacing of the energy modulation and the bunch length to energy chirp calibration, we estimate that the frequency of the TPIPE mode is ($459 \pm 32$)~GHz, taking into account the resolution of the energy spectrometer and the uncorrelated beam energy spread. From the energy modulation measurement we estimate the maximum decelerating field inside the long bunch is ($1 \pm 0.2$)~MV/m. This field value is a measure of the bunch current rise time.

\subsection*{Interferometer Measurements}

\begin{figure}[htb]
\centering
\includegraphics*[width=68mm]{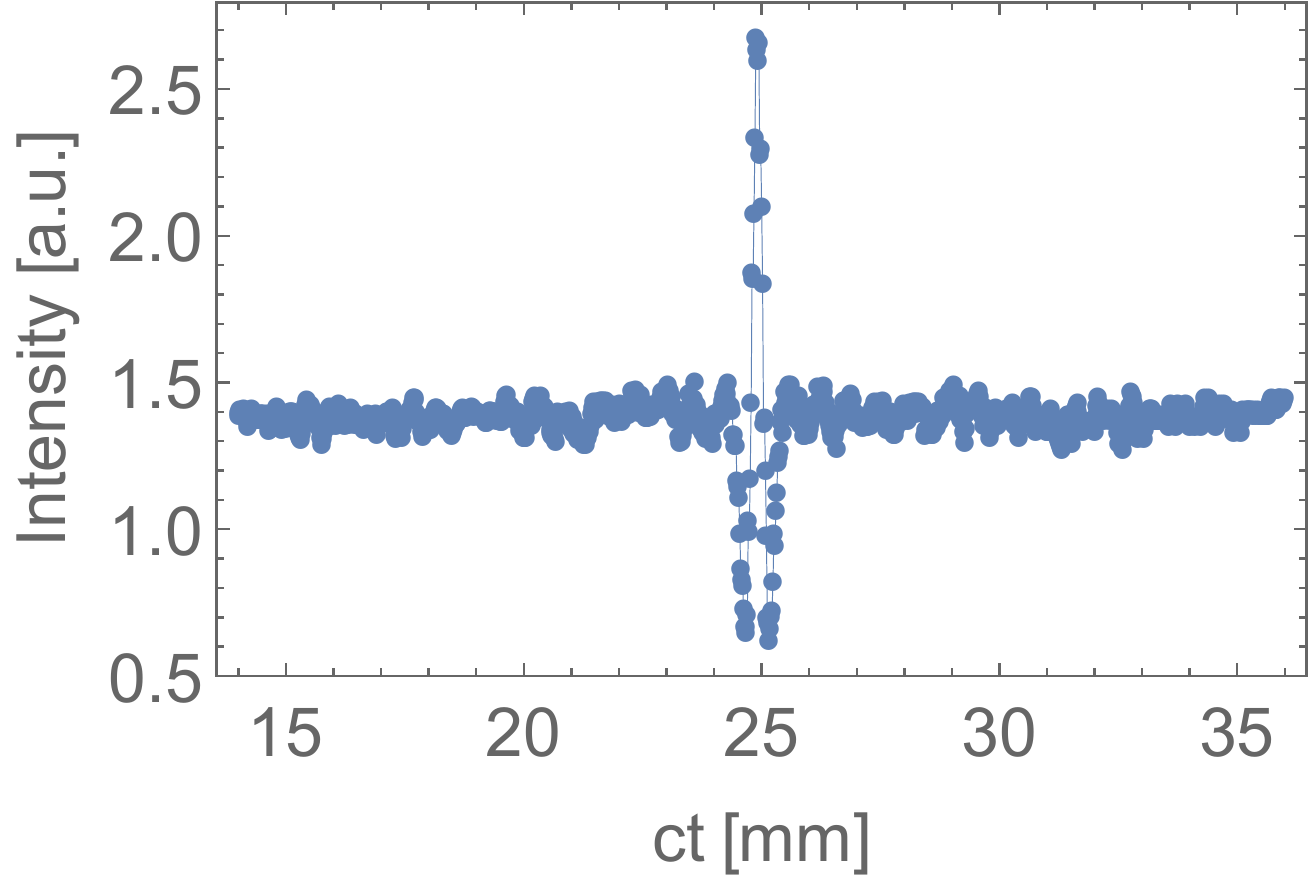}
\includegraphics*[width=68mm]{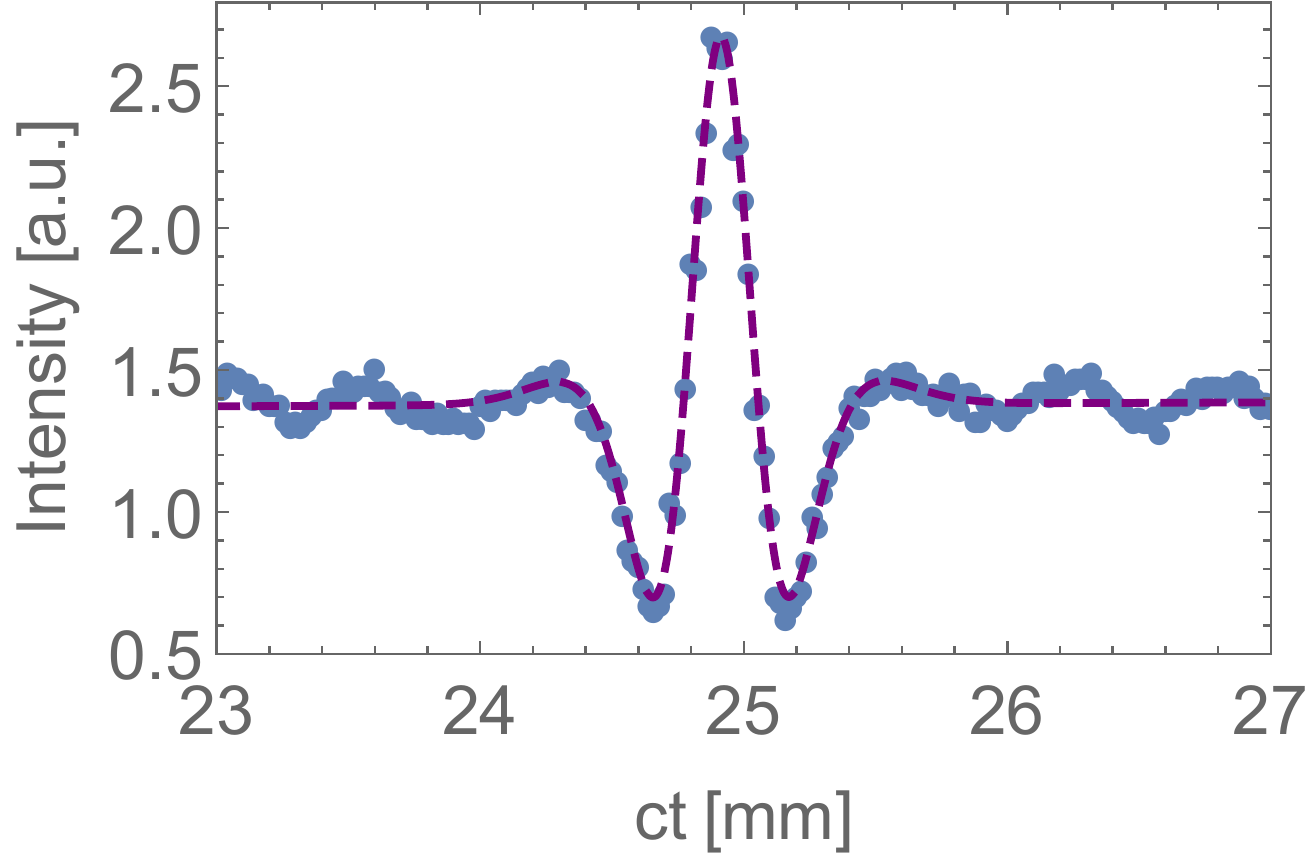}
\caption{(Left frame) A fine, 22-mm interferometer scan of the radiation generated by TPIPE; $ct$ is path length change of the light in one arm of the interferometer (with arbitrary offset).
The data points (the symbols) are connected by straight lines to aid the eye. (Right frame) A zoomed-in shot of the center of the scan. Here the purple, dashed curve gives the best fit to the diffraction radiation model for a Gaussian beam, discussed below.}
\label{scantpipe_fi}
\end{figure}

For the interferometer scans we need a shorter bunch, generated using a mask, as described above.
A 22-mm-long interferometer scan (with a 20~$\mu$m step size), with the beam passing through TPIPE, is shown in Fig.~\ref{scantpipe_fi} in the left plot, where $ct$ gives the path length change in the light in one arm of the interferometer, equal to twice the distance of travel of the movable mirror in the interferometer (both with arbitrary offset). This plot is actually the composite of three scans, where we verified that the oscillations at the boundaries (of the three scans) matched well in amplitude and phase of oscillation. The bunch charge was $q=50$~pC. The right plot in Fig.~\ref{scantpipe_fi} zooms in to show more details of the central portion of the scan.

\begin{figure}[htb]
\centering
\includegraphics*[width=88mm]{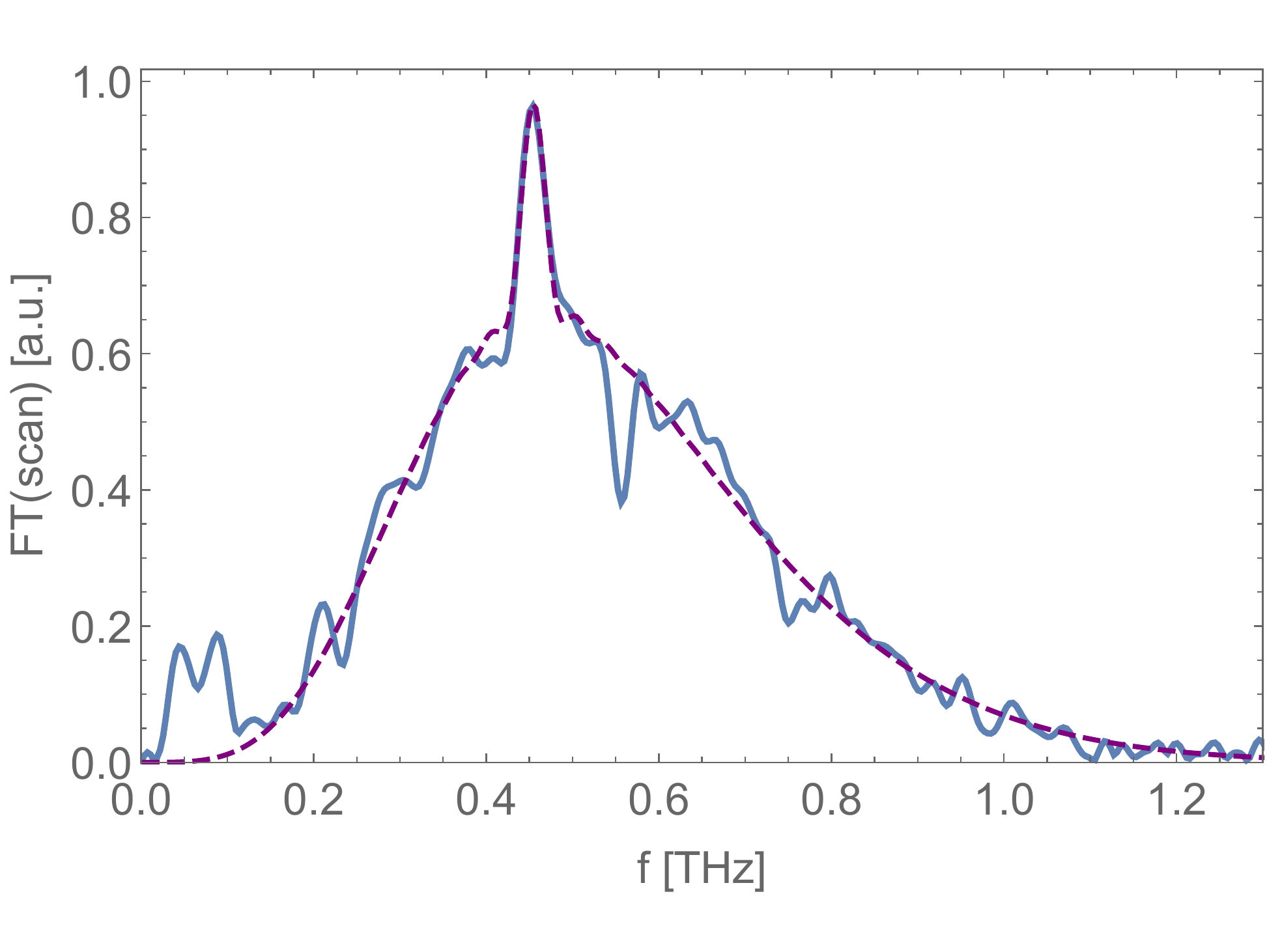}
\caption{The absolute value of the Fourier transform of the scan of Fig.~\ref{scantpipe_fi} (the blue, solid curve). The dashed, purple curve gives the best model fit. }
\label{scantpipeb_fi}
\end{figure}

After padding with zeros,
% (for a more detailed spectrum plot), 
a Fast Fourier Transform (FFT) of the scan of Fig.~\ref{scantpipe_fi} (on the left) was performed. The absolute value of the Fourier transform is given in Fig.~\ref{scantpipeb_fi}. We see a rather broad peak topped by a narrow horn. We consider this spectrum to be the sum of the spectra of: the narrow-band THz pulse generated at TPIPE (the signal), and the broad-band, diffraction radiation pulse generated at the mirror (the background). %We justify this by the fact the the two signals are (mostly) separated by time, with the diffraction radiation pulse arriving before the THz pulse.
%in the following way:
%We know that fields from two sources add, due to the principle of superposition. With two sources of energy (like what is obtained by the interferometer), however, there can, in general, be interference. In our case, however, since the two sources are separated in time---the diffraction radiation pulse arrives first at the mirror---there can be no interference and linear superposition also applies. 

The broad spectrum that is recorded has a low frequency cut-off given by the apertures in the interferometer, and a high frequency cut-off determined by the bunch spectrum. There are features in the plot that are not understood and that we consider artifacts, such as a peak near $f=100$~GHz and some ripples in the spectrum. In our experiment, no measures were taken to keep water vapor out of the interferometer, and prominent dips seen in the spectrum at $f=557.1$~GHz, 750.5~GHz, closely match the first two absorption frequencies of water vapor, 557.~GHz and 752.~GHz (see {\it e.g.} \cite{Slocum}; these dips were found in all our scans). These dips can serve as indications of the accuracy of the frequency values in the plot.

To approximate the broad, diffraction radiation background we choose the function~\cite{Leissner99}
\begin{equation}
\tilde S(k,\sigma_z,\zeta)=e^{-k^2\sigma_z^2}(1-e^{-k^2 \zeta^2})^2\ ,\label{fit_eq}
\end{equation}
with the two fitting parameters bunch length $\sigma_z$ and $\zeta$; the high frequency behavior is that generated by a Gaussian bunch with rms length $\sigma_z$. Note that, in the time domain, this function becomes~\cite{Leissner99} 
\begin{equation}
S(s,\sigma_z,\zeta)= \left(e^{-\frac{z^2}{4/\sigma_z^2}}
-\frac{2\sigma_z\cdot e^{-\frac{s^2}{4(\sigma_z^2+\zeta^2)}}}{\sqrt{\sigma_z^2+\zeta^2}}
+\frac{\sigma_z\cdot e^{-\frac{s^2}{4(\sigma_z^2+2\zeta^2)}}}{\sqrt{\sigma_z^2+2\zeta^2}} \right)\ ,\label{fitt_eq}
\end{equation}
with $s=ct$.

As our model spectrum function, we take the sum of the diffraction radiation background (Eq.~\ref{fit_eq}) and the model of the THz signal ($|\tilde E_r|^2$ taken from Eq.~\ref{Etilde_rect_eq}); specifically, we take intensity to be given by 
\begin{equation}
I(f)=\alpha_1 {\tilde S}(2\pi f/c,\sigma_z,\zeta) + \alpha_2 |\tilde E_r(f,f_c,n_\lambda)/E_{r0}|^2\ ,\label{fitb_eq}
\end{equation}
with fitting parameters $\alpha_1$, $\sigma_z$, $\zeta$, for the broad-band diffraction radiation background, and $\alpha_2$, $f_c$, $n_\lambda$, for the narrow-band THz signal. 
Before fitting to the spectrum of Fig.~\ref{scantpipeb_fi}, we remove the artificial peak near $f=100$~GHz and the dip due to water vapor at $f=557$~GHz. We fit using the { Maximum-Likelihood Method} (see {\it e.g.} \cite{BevingtonRobinson}), where it is assumed that the measured values of $I(f)$ have independent, normally distributed errors. This method, in addition to finding the best fitting parameters, gives us also an estimate of their errors.

The best fit is shown by the dashed, purple curve in Fig.~\ref{scantpipeb_fi}. We see that the fit is indeed good.  For the parameters connected to diffraction radiation, we find that the fitted $\sigma_z=(87.\pm1)$~$\mu$m and $\zeta=(137.\pm2)$~$\mu$m. That the fit is quite good at the higher frequencies indicates that a Gaussian bunch of $\sigma_z=87$~$\mu$m gives a reasonable approximation to the bunch shape. 
%In principle, we could take the spectral shape and, using Kramers-Kronig relations, find an approximation to the actual bunch shape (see, {\it e.g.} Ref.~xxx); however, we believe knowing the details of the bunch shape is not important for the purpose of this report and will forgo this exercise.
In the time domain, the fitted filter function, Eq.~\ref{fitt_eq}, also matches the central part of the data well (see Fig.~\ref{scantpipe_fi} on the right, the purple dashed curve).

The narrow peak in the spectrum of Fig.~\ref{scantpipeb_fi}, corresponding to the THz signal, is also fit well by the model. The relevant fitted parameters are: $f_c=(454.2\pm0.6)$~GHz and $n_\lambda=13\pm\frac{1}{2}$. The fitted parameters imply (see Eq.~\ref{ell_eq}): the quality factor, $Q=10.8 \pm 0.4$, and the pulse length, $\ell=(8.6 \pm 0.3)$~mm.
The central frequency agrees with the spectrometer measurement discussed above and (reasonably well) with the ECHO calculations (471~GHz). The 4\% disagreement with the calculations suggests that there is still some error in our understanding of the shape of the corrugations. %For example, the scaling of the central frequency $f_c$ with depth of corrugation $\delta$ for the case of rectangular corrugations ($f\sim\delta^{-1/2}$), suggests that the actual depth of corrugation may be $\sim10\%$ larger than we think. 
%Note that a 10\% change in depth of corrugation, for example, would be enough to remove the disagreement.

As for the $Q$, note that there is a maximum value of $Q$ that can be resolved from an interferometer scan.  An interferometer scan, in essence, is the result---after high-pass filtering of the radiation pulse---of performing an autocorrelation. For a pulse of finite length, the autocorrelation will have double that length. The interferometer scans will have a finite total difference in path length of the light; let us denote it by $\Delta ct$. From the discussion in the ``Bandwidth of THz Pulse" section above and from Eq.~\ref{ell_eq}, we find that for given $\Delta ct$, the maximum quality factor that can be resolved is 
\begin{equation}
Q_{max}=\frac{f_c\Delta ct}{2.4c}\ .\label{Qmax_eq}
\end{equation}
%For example, for $f_c=450$~GHz and $\Delta ct= 22$~mm (for a scan that will be analyzed below), $Q_{max}=14$.
Here $Q_{max}=14$; thus, it appears that our fitted $Q$ (10.8) is not limited by the resolution of the measurement.

The measured ratio, at $f_c=454$~GHz, of the spectral power of the THz signal to that of the diffraction radiation background  is $\Lambda=0.50 \pm 0.05$. The calculated result, given above ($\Lambda=0.30$), is only 60\% of this value, 
%The fact that the measured result is larger than the calculated one 
suggesting that the amount of diffraction radiation that was recorded is less than expected. We originally thought that one reason this might be is that---between the exit of TPIPE and the location of the mirror---the beam's primary fields have not had a chance to reconstitute themselves; a simulation with ECHO, however, ruled out this as a significant effect here. Another possible reason for a disagreement in the diffraction radiation may be due to the fact that the mirror geometry is not simply a flat, round disc orientated perpendicular to the beam trajectory; this could be verified by {\it e.g.} performing a 3D ECHO simulation with the real geometry. Nevertheless, given the simplicity of the model in the calculations, we believe that an agreement to 60\% represents reasonably good agreement. A summary of the main results obtained from the spectrum of Fig.~\ref{scantpipeb_fi} can be found in Table~\ref{results_tab}. 

\begin{table}[h!]
\centering
\caption{Summary of main results obtained from interferometry: properties of the THz pulse, comparing calculations with analysis of the scan of Fig.~\ref{scantpipe_fi}. Note that the experimental value of pulse length $\ell$ is estimated from the measured quality factor $Q$ according to Eq.~\ref{ell_eq}. 
%Error values given mean error estimates of model fit to measurement.
}
\begin{tabular}{||l|c|c|c ||} \hline\hline
Parameter&Theory &Experiment & Units\\ \hline\hline 
Central frequency, $f_c$&471. &454.2 &GHz\\ \hline
Quality factor, $Q$&8.3 &10.8  & \\ \hline
Pulse length (FWHM), $\ell$&5.8 &8.6 &mm\\ \hline
Relative strength, $(dU/df)$, at $f=f_c$, $\Lambda$&0.30 & 0.50 &\\ \hline\hline
\end{tabular}\label{results_tab}
\end{table}

\begin{figure}[htb]
\centering
\includegraphics*[width=70mm]{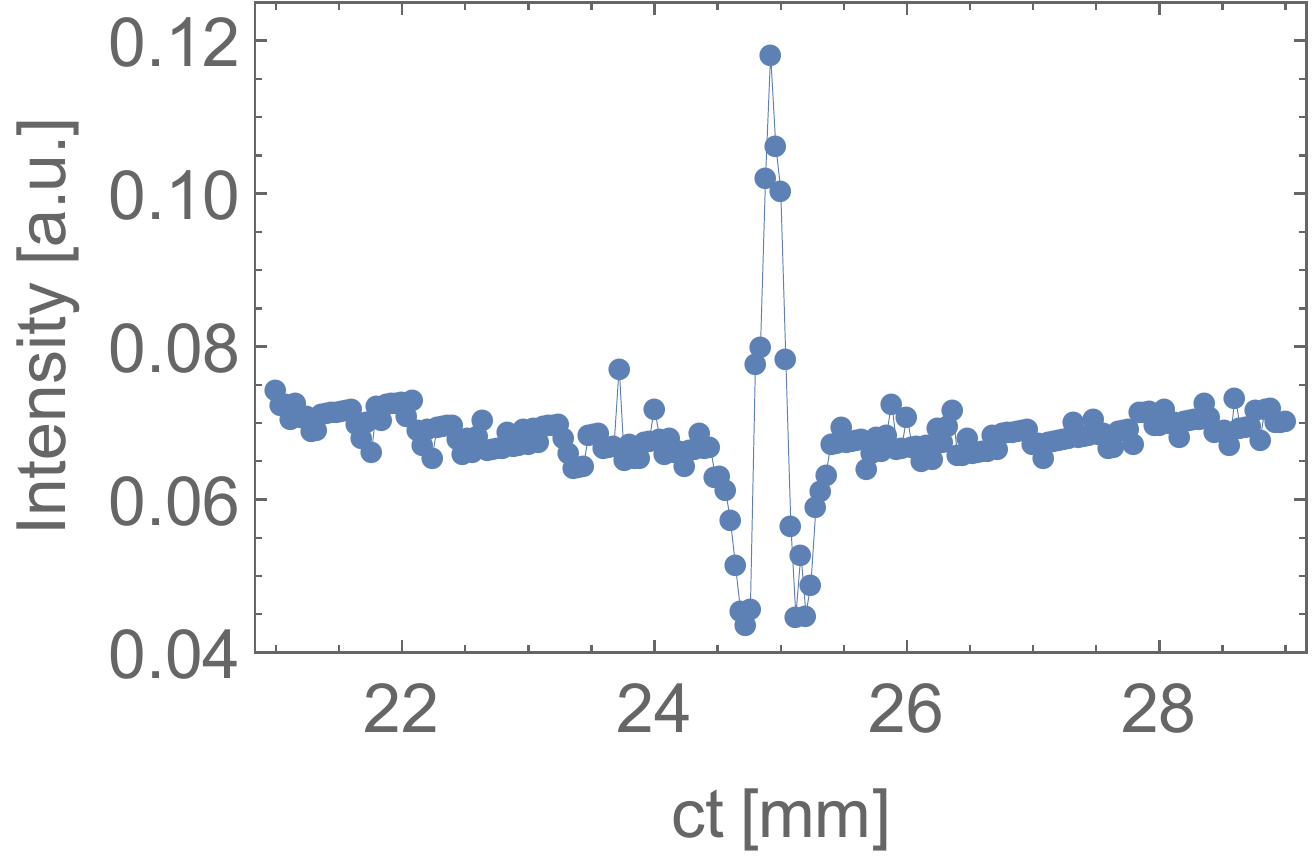}
\includegraphics*[width=68mm]{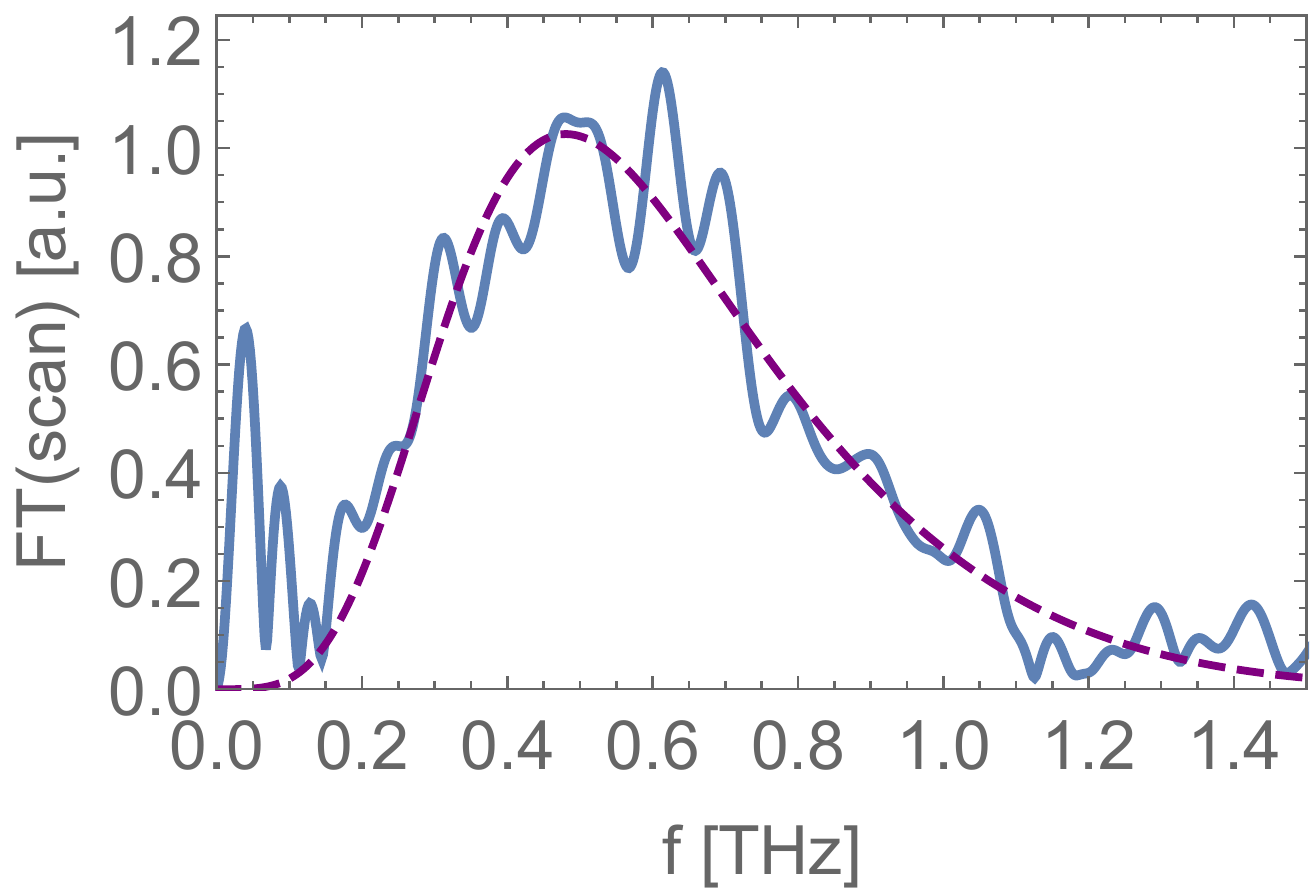}
\caption{Interferometry scan and spectrum for the case of the beam passing through the smooth bore of TPIPE.}
\label{scansmooth_fi}
\end{figure}

\begin{figure}[htb]
\centering
\includegraphics*[width=70mm]{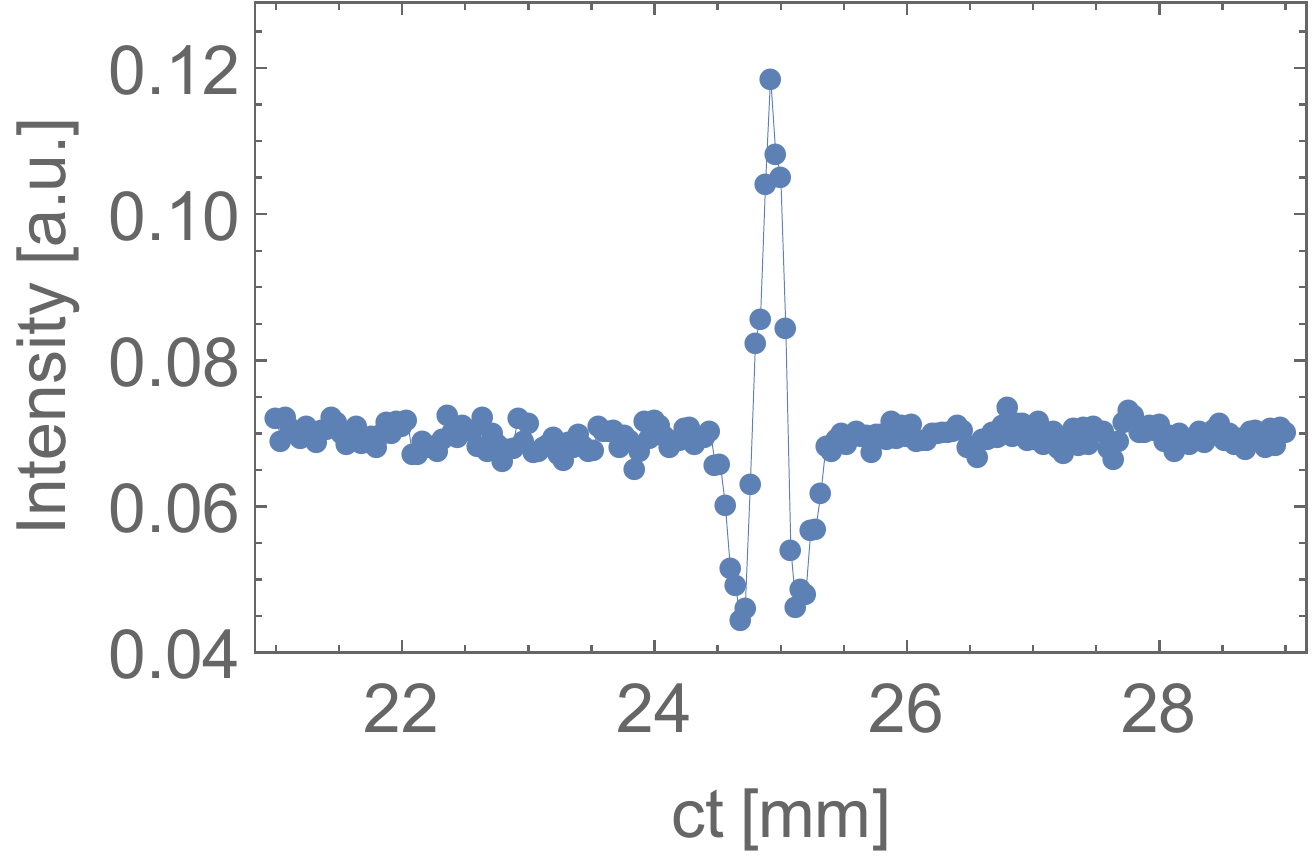}
\includegraphics*[width=68mm]{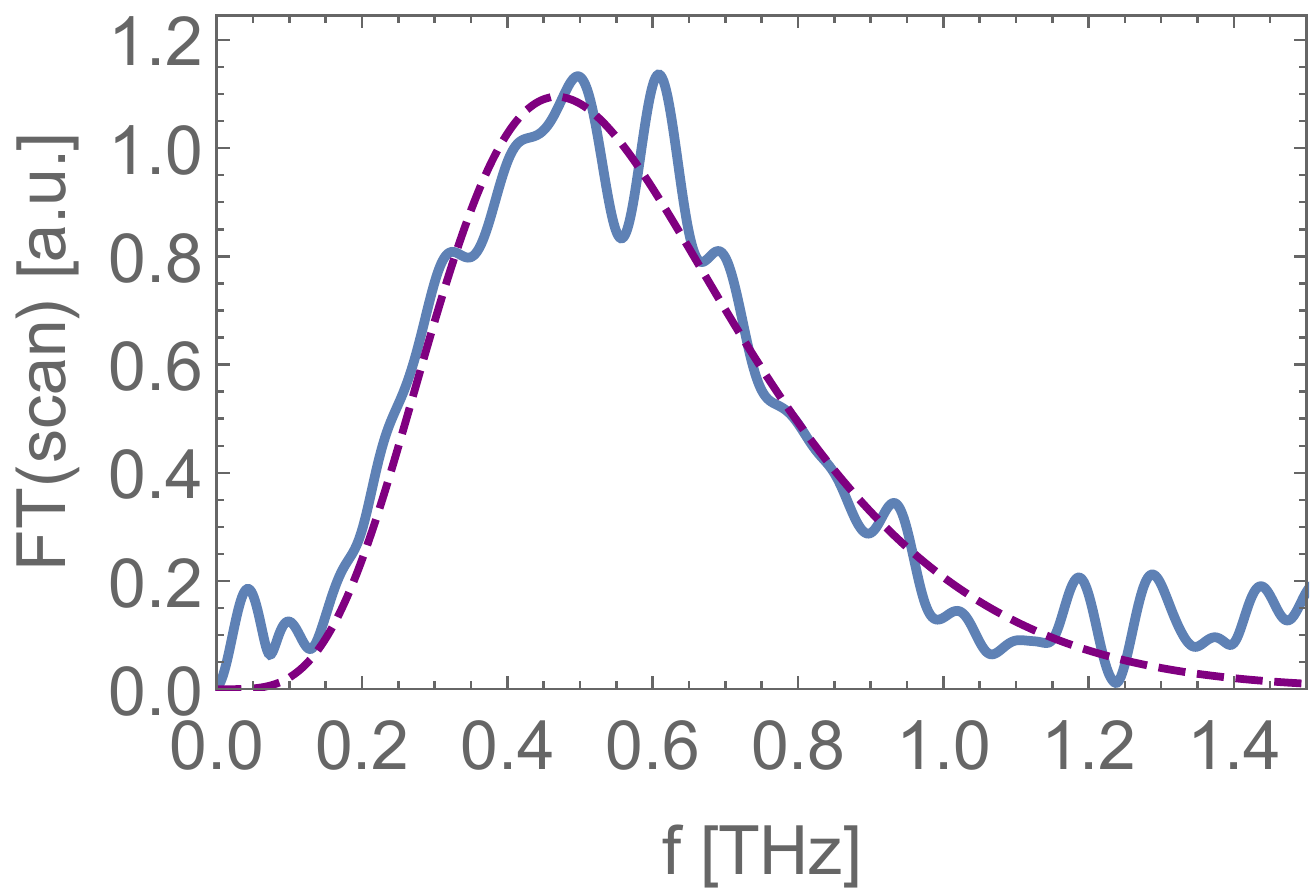}
\caption{Interferometry scan and spectrum for the case of TPIPE withdrawn from the path of the beam.}
\label{scannone_fi}
\end{figure}

For the null test, we present in Figs.~\ref{scansmooth_fi}--\ref{scannone_fi},  interferometer results for the case of the beam passing through the smooth tube in TPIPE, and the case TPIPE is withdrawn completely from the path of the beam  (left plots gives the scan, the right ones the Fourier transform).
%; the results for TPIPE removed from the path of the beam are similar). 
These scans were performed one after the other late in the data taking, when the helium was nearly depleted; this fact resulted in a much weaker signal and probably accounts for the larger number of artifacts (oscillations) seen in the spectrum (especially the smooth pipe case). In addition, the scans only included a total path length difference, $\Delta ct=8$~mm, reducing the resolution in frequency. We again see the dips near $f=555$~GHz, 750~GHz, but this time there is no clear horn or signal on top that can be identified as a narrow-band THz signal. The spectra, in both cases, are essentially the same. The dashed, purple curve in the plots give the fit to the function,
Eq.~\ref{fit_eq}; in Fig.~\ref{scansmooth_fi} with $\sigma_z=68$~$\mu$m and $\zeta=155$~$\mu$m; in Fig.~\ref{scannone_fi} with $\sigma_z=74$~$\mu$m and $\zeta=153$~$\mu$m.

We had more interferometer scans with TPIPE, and for different bunch lengths. However, many were of poor quality, and most were for total scan length $\Delta ct=4$~mm, where there is insufficient resolution for obtaining $f_c$ or $\Lambda$, and certainly not for $Q$. One good measurement with $\Delta ct=8$~mm yielded a spectral peak at $f_c=456.2$~GHz, in reasonable agreement with the value of 454.2~GHz obtained by our best scan, presented in detail above.

\section*{Conclusions}

We have demonstrated the generation of a narrow-band THz pulse by having a relativistic beam pass through a round, corrugated metallic pipe by two methods: (1)~using a relatively long bunch (with short rise time) to energy modulate the beam, that---after introducing an energy chirp---was detected in a spectrometer; (2)~using a short bunch to generate a narrow-band THz pulse that was detected using an interferometer and a LHe bolometer. The null cases---those of a smooth pipe or with the corrugated structure completely removed from the beam path---found no narrow-band signal.

In addition to this proof-of-principle demonstration, we measured (and compared with calculations): the central frequency of the pulse, $f_c=454$~GHz, the effective quality factor $Q=10.8$, and the relative strength of the signal at the central frequency (to a diffraction radiation background), $\Lambda=0.50$. These results are all in reasonable agreement with our calculations. From the $Q$ value, we can estimate the THz pulse full length to be $\ell=8.6$~mm.

We have only one good measurement of TPIPE with the interferometer, and thus have no good estimate of the experimental errors. Considering the central frequency $f_c$, for example, one can conservatively take the error estimate to be $c/\Delta ct=15$~GHz ($\Delta ct$ is the full scan range of path length difference). However, comparing prominent dips in the spectrum to the first two absorption lines of water vapor and finding differences $\le1.5$~GHz, suggest that the accuracy is much better. 

\section*{Acknowledgments}
We thank: Makino Machine Tools for machining TPIPE for us free of charge; G. Bowden, the engineer on the TPIPE project, for his careful work; the Accelerator Test Facility staff at BNL for engineering support during the experiment;  the UCLA PBPL group for letting us use their interferometer set-up; A. Fisher for helpful discussions, drawing on his experience in interferometry analysis. Euclid Beamlabs LLC acknowledges support from US DOE SBIR program grant No. DE-SC0009571. 
 Work was partially supported by Department of Energy contract DE--AC02--76SF00515.

\appendix
\section{Fraction of THz Energy that Reaches the Mirror}

We compute here the relative loss of spectral power in the THz pulse---between the exit of the corrugated structure and the arrival at the mirror---due to diffraction of the wave. The electric field at the exit of the corrugated pipe is approximated by
    \begin{align}\label{eq:1}
    \vec{\mathcal {E}}(\vec r)
    =
    A
    {\vec r}
    e^{-2\pi if t}
   \  ,
    \end{align}
with $A$ a constant, $\vec r$ is the transverse position, $f$ is the frequency, and $t$ is time. (We will drop the $e^{-2\pi i ft}$ in the following equations; this factor does not affect spectral energy results.)
The spectral energy of this field
    \begin{align}\label{eq:2}
    W
    =\frac{1}{4\pi}
    \int
    2\pi r dr\, 
    |\vec{\mathcal {E}}(\vec r)|^{2}
    =
    \frac{1}{2}
    A^{2}
    \int_{0}^{a}
    r^{3}dr
    =
    \frac{1}{8}
    A^{2}
    a^{4}\ ,
    \end{align}
where $a$ is the radius of the corrugated pipe.

To find the distribution of the electric field at the downstream mirror, $\vec E_{\mathrm{m}}(\vec r')$, we use vectorial diffraction theory~\cite{Jackson}
    \begin{align}\label{eq:3}
    {{\vec E}}_\mathrm{m}(\vec r')
    =
    -
    e^{ikL+ ik{r'}^{2}/2L}
    \frac{ik}{2\pi L}
    \int
    e^{{-ik\vec{r'}\cdot\vec{r}/L}}
    \vec{\mathcal {E}}(\vec r)\,
    d^2r
   \ ,
    \end{align}
where $k=2\pi f/c$, $L$ is the distance between corrugated pipe and mirror, and the integral is performed over the cross-section of the pipe. It is clear that $\vec E_{\mathrm{m}}$ is directed along $\vec{r}'$; {\it i.e.}  ${{\vec E}}_\mathrm{m}(\vec r') = E_\mathrm{m}(r')\vec{r}'/r'$. For $E_{\mathrm{m}}$ we obtain
    \begin{align}\label{eq:4}
    E_\mathrm{m}
    &=
    e^{ikL+ ik{r'}^{2}/2L}
    \frac{ik}{2\pi L}
    A
    \int
    d^2 r\,
    e^{{-ik\vec{r'}\cdot\vec{r}/L}}
    \vec{r}\cdot
    \frac{\vec{r}'}{r'}
    \nonumber\\
    &=
    e^{ikL+ ik{r'}^{2}/2L}
    \frac{ik}{2\pi L}
    A
    \int_0^\infty
    r^{2}dr 
    \int_0^{2\pi}
    d\phi\,
    e^{{-ik{r'}{r}\cos\phi /L}}
    \cos\phi
   \  .
    \end{align}
The integral over $\phi$ can be taken using 
	\begin{align}\label{eq:5}
	\int_0^{2\pi}e^{-i\alpha\cos\phi}\cos\phi 
	\,d\phi
	=
	-2\pi iJ_1(\alpha)\ ,
	\end{align}
with the result
    \begin{align}\label{eq:6}
    E_\mathrm{m}
    &=
    e^{ikL+ ik{r'}^{2}/2L}
    \frac{k}{ L}
    A
    \int_0^a
    r^{2}dr \,
    J_1\left(\frac{k{r'}{r}}{L}\right)
    \nonumber\\
    &=
    e^{ikL+ ik{r'}^{2}/2L}
    A
    \frac{a^{2}}{r'}
    J_2\left(\frac{k{r'}a}{L}\right)
    .
    \end{align}

We compute the spectral energy reaching the mirror by using~\eqref{eq:2} with the electric field at the mirror~\eqref{eq:6}
    \begin{align}\label{eq:7}
    W_\mathrm{m}
    &=\frac{1}{4\pi}
    \int_{0}^{b}
    2\pi r' dr'
    |E_\mathrm{m}(r')|^{2}
    =
    \frac{1}{2}
    A^{2}
    a^4
    \int_{0}^{b}
    \frac{dr'}{r'}
    J_2^{2}\left(\frac{k{r'}a}{L}\right)
    \nonumber\\
    &=
    \frac{1}{8}
    A^{2}
    a^4
    F\left(\frac{kab}{L}\right)
    ,
    \end{align}
where $b$ is the mirror radius and
    \begin{align}\label{eq:8}
    F(x)
    =
    1
    -
    \left(\frac{4}{x^2}+2\right) J_1(x){}^2
    -2
    J_0(x){}^2
    +
    \frac{4 }{x}J_1(x)
    J_0(x)\ .
    \end{align}
Finally, the  ratio of the THz spectral power that arrives at the mirror to that at the exit of the corrugated pipe is
    \begin{align}\label{e,q:9}
    \rho_{\mathrm{THz}}\equiv\frac{W_\mathrm{m}}{W}
    =
    F\left(\frac{kab}{L}\right)
   \ .
    \end{align}

%%%%%%%%%%%%%%%%%%%%%%%%%%%%%%%%%%%%%%%%%%%%%%%

\section{Diffraction Radiation Background}

An ultra-relativistic beam passes through a hole in a mirror. The electromagnetic energy hitting the mirror can be written in terms of the radial electric field $E(r,z)$ as 
    \begin{align}\label{eq:11}
    U
    =
    \frac{1}{4\pi}
    \int_{-\infty}^{\infty} 
    dz
    \int_{b_{1}}^{b}
    2\pi r dr\,
    E(r,z)^{2}
   \  ,
    \end{align}
(we took into account that the magnetic field is equal to the electric one). The lower limit $b_{1}$ is the radius of the hole in the mirror. If the bunch longitudinal distribution function $\lambda(z)$ is normalized to unity, then $E(r,z)$ can be computed as a superposition of the field of a point charge $E_{p}(r,z)$: 
    \begin{align}\label{eq:12}
    E(r,z)
    =   
    q 
    \int
    dz' \lambda(z')
    E_{p}(r,z-z')
   \ ,
    \end{align}
where $q$ is the total charge of the beam. 

Fourier transforming \eqref{eq:12}, we find
    \begin{align}\label{eq:13}
     \tilde E(r,k)
    =
    \int_{-\infty}^{\infty} 
    dz\,
    e^{-ikz}E(r,z)
    =q
      \tilde \lambda(k)
     \tilde E_{p}(r,k)\ .
    \end{align}
From the Parseval's theorem
    \begin{align}\label{eq:14}
    \int_{-\infty}^{\infty} 
    dz\,
    E(r,z)^{2}
    =
    \frac{1}{\pi}
    \int_{0}^{\infty}
    dk\,
    | \tilde E(r,k)|^{2}
    \ ;
    \end{align}
hence the spectral energy (with dimensions J/Hz)
    \begin{align}\label{eq:15}
    \left(\frac{dU}{df}\right)_{\mathrm d}
    =
    \frac{1}{ c}
    \int_{b_{1}}^{b}
    r dr\,
    | \tilde E(r,k)|^{2}
    =
    \frac{q^{2}| \tilde \lambda(k)|^{2}}{ c}
    \int_{b_{1}}^{b}
    r dr\,
    | \tilde E_{p}(r,k)|^{2}
    ,
    \end{align}
where $f=ck/2\pi$. The Fourier transform of the field of a relativistic point charge is well known,
    \begin{align}\label{eq:16}
    | \tilde E_{p}(r,k)|
    =
	\frac{2k}{\gamma}
	K_{1}\left(\frac{kr}{\gamma}\right)\ .
    \end{align}
Thus, we obtain
    \begin{align}\label{eq:17}
    \left(\frac{dU}{df}\right)_{\mathrm d}
    =
    \frac{4q^{2}| \tilde \lambda(k)|^{2}}{c}
    \int_{b_{1}}^{b}
    r dr
	\frac{k^{2}}{\gamma^{2}}
	K_{1}\left(\frac{kr}{\gamma}\right)^{2}
    =
    \frac{4q^{2}| \tilde \lambda(k)|^{2}}{c}
    \left[
	G\left(\frac{kb}{\gamma}\right)
	-
	G\left(\frac{kb_{1}}{\gamma}\right)
	\right]
   \  ,
    \end{align}
with
    \begin{align}\label{eq:18}
    G(x)
    =
    \frac{1}{2} x^2
   \left[K_1(x){}^2-K_0(x)
   K_2(x)\right]
   .
    \end{align}

%\begin{thebibliography}{99}   % Use for  10-99  references

\end{document}